%% file: lattice12.tex
\documentclass{PoS}

\usepackage{bm}% bold math
\usepackage{amsmath}
\usepackage[square,comma,numbers,sort&compress]{natbib}
\usepackage{color}
\usepackage{graphicx}
\usepackage{bbm}

\usepackage{amsfonts}

\usepackage{pst-grad}
\usepackage{dsfont}
\usepackage{tikz}
\usetikzlibrary{shapes,arrows,backgrounds}
\usepackage{placeins}

\newcommand{\rep}{\mathcal{R}}

\newcommand*{\no}{\noindent}
\newcommand*{\bea}{\begin{eqnarray}}
\newcommand*{\eea}{\end{eqnarray}}
\newcommand*{\be}{\begin{equation}}
\newcommand*{\ee}{\end{equation}}

\newcommand*{\nn}{\nonumber}

\newcommand{\includeEPSTEX}[1]{\includegraphics{#1}}
\graphicspath{{./}{publPlots/}}

\title{G$_2$ gauge theories}

\ShortTitle{G$_2$ gauge theories}

\author{\speaker{Axel Maas}\thanks{Supported by the DFG under grant number MA 3935/5-1.}\\
        E-mail: \email{axelmaas@web.de}}

\author{\speaker{Bj\"orn H.\ Wellegehausen}\thanks{Supported by the DFG graduate school 1523-1.}\\
        E-mail: \email{bjoern.wellegehausen@uni-jena.de}\\

        Institute of Theoretical Physics, Friedrich-Schiller-University Jena, Max-Wien-Platz 1, D-07743 Jena, Germany}

\abstract{QCD can be formulated using any gauge group. One particular interesting choice is to replace SU(3) by the exceptional group G$_2$. Conceptually, this group is the simplest group with a trivial center. It thus permits to study the conjectured relevance of center degrees of freedom for QCD. Practically, since all its representation are real, it is possible to perform lattice simulations for this theory also at finite baryon densities. It is thus an excellent environment to test methods and to investigate general properties of gauge theories at finite densities. We review the status of our understanding of gauge theories with the gauge group G$_2$, including Yang-Mills theory, Yang-Mills-Higgs theory, and QCD both in the vacuum and in the phase diagram.}

\FullConference{The 30 International Symposium on Lattice Field Theory - Lattice 2012,\\
		June 24-29, 2012\\
		Cairns, Australia}

\begin{document}

\section{Introduction}

Gauge theories can in principle, up to anomalies, be formulated for all simple Lie groups. This property has been used often to gain insight into structures or simplify calculations. One salient example is the large-$N$ limit in QCD. Another option is to use the exceptional group G$_2$, leading to G$_2$ QCD.

The proposal to make this replacement was made \cite{Holland:2003jy} to understand the role of the center of the gauge group, which was long assumed to play a central role for many of the salient features of QCD, especially confinement. However, the detailed investigations, to be presented in section \ref{sym}, showed that most of these features are also present in the G$_2$ case.

Besides these conceptual questions concerning the center, another property of G$_2$ QCD is interesting from a practical point of view. Since all its representations are real, no sign problem arises when simulating G$_2$ QCD with dynamical fermions. It is thus possible to investigate the whole phase diagram of the theory using lattice calculations \cite{Maas:2012wr}. G$_2$ QCD is so far the theory most similar to QCD where this is possible in the continuum limit. The resulting phase diagram \cite{Maas:2012wr} is rather similar to the one obtained in other such theories, like QCD with gauge group SU(2) (QC$_2$D) \cite{Hands:2006ve,Hands:2011ye,Strodthoff:2011tz,Cotter:2012tt} or QCD in the strong coupling limit \cite{deForcrand:2009dh,Fromm:2011zz}. Thus, G$_2$ QCD offers another perspective on the QCD phase diagram. This will be detailed in sections \ref{sqcd} and \ref{ssmall}.

It is, of course, an interesting question whether there can be established any direct connection between the G$_2$ case and the SU(3) world. Breaking the G$_2$ gauge group using a Higgs field works for the Yang-Mills case \cite{Holland:2003jy}, as briefly outlined in section \ref{shiggs}, but it is yet not clear whether this is also possible in the QCD case.

Thus, gauge theories with gauge group G$_2$ are very interesting from many perspectives, as will be summarized in section \ref{sconclusion}. However, most investigations are yet on a qualitative and exploratory level, and many interesting questions have not even been addressed yet.

\section{Yang-Mills theory}\label{sym}

\subsection{Zero temperature}

The simplest realization of a gauge theory with the gauge group G$_2$ is Yang-Mills theory. Since the adjoint representation of G$_2$ is 14-dimensional, there are 14 gluons. Using the Macfarlane representation \cite{Macfarlane:2002hr} a G$_2$ link (or group element) $U$ in the 7-dimensional fundamental representation can be written as
\be
U=Z\begin{pmatrix} u & 0 & 0 \cr 0 & 1 & 0 \cr 0 & 0 & u^* \end{pmatrix},\nn
\ee
\no where $Z$ is a 7-dimensional representation of $S^6$ and $u$ is an element of SU(3). Thus, 8 of the gluons can be considered loosely as 'SU(3)'-like. This will become important in section \ref{shiggs}. Due to this explicit SU(3) subgroup, lattice simulations of a G$_2$ theory are straightforward but expensive, see \cite{Maas:2007af,Wellegehausen:2011sc,Maas:2012wr,Wellegehausen:2010ai} for the algorithms employed here.

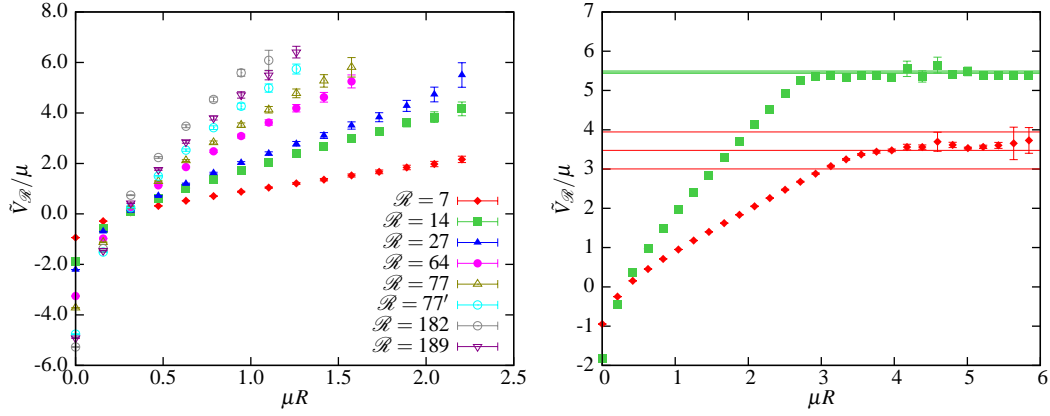
\begin{figure}
\input{potential40}\input{StringBreakingPot}
\caption{\label{fig:pot}The Wilson potential $\tilde{V}$ divide by the scale $\mu$ for different representations $\rep$ (left) and its string-breaking, compared to hybrid masses for two representations (right), both in three dimensions, from \cite{Wellegehausen:2010ai}.}
\end{figure}

G$_2$ is the smallest rank 2 gauge group with a trivial center. As a consequence, every fundamental charge can be screened by three adjoint charges, and thus there is no infinitely rising Wilson potential, and thus no confinement in the sense of a Wilson area law \cite{Holland:2003jy}. However, in practice the corresponding Polyakov loop is found to be very small at zero temperature, and in fact only upper bounds are known, though it follows that it must be non-zero. In fact, at intermediate distances a linear rising Wilson potential \cite{Pepe:2006er,Greensite:2006sm}, including a characteristic Casimir scaling \cite{Wellegehausen:2010ai,Liptak:2008gx}, is found. Thus, a string appears in the same way as in QCD with dynamical quarks, up to a distance where string-breaking sets in \cite{Wellegehausen:2010ai}. Hence, G$_2$ Yang-Mills theory is in the same sense (non-)confining as is QCD. These facts are illustrated in figure \ref{fig:pot}. Of course, since the theory has no anomaly, it is still a well-defined theory, with only colorless asymptotic states \cite{Holland:2003jy,Pepe:2006er}, like glueballs \cite{Wellegehausen:2010ai,Lacroix:2012pt}.

\begin{figure}
\includegraphics[width=0.5\linewidth]{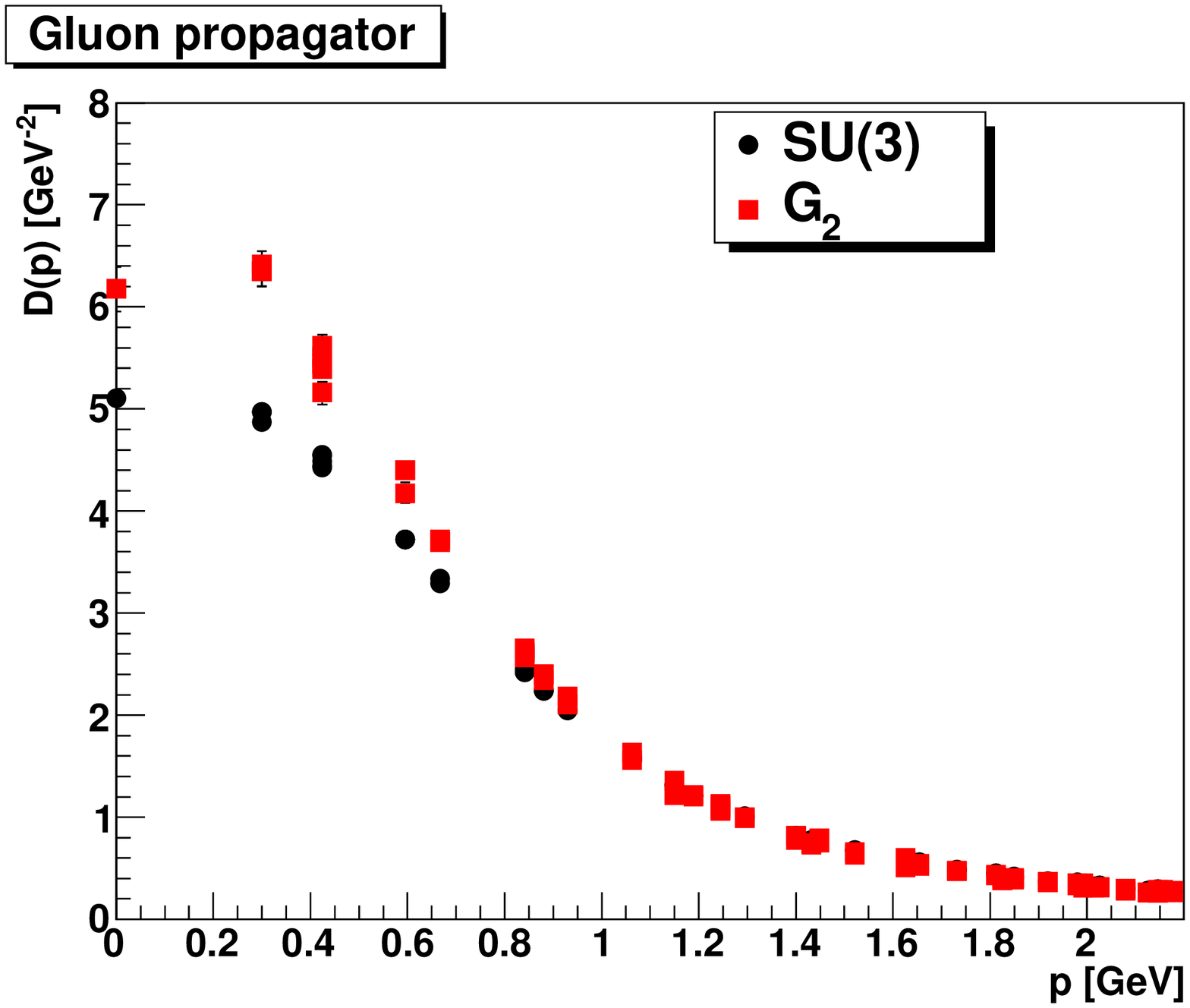}\includegraphics[width=0.5\linewidth]{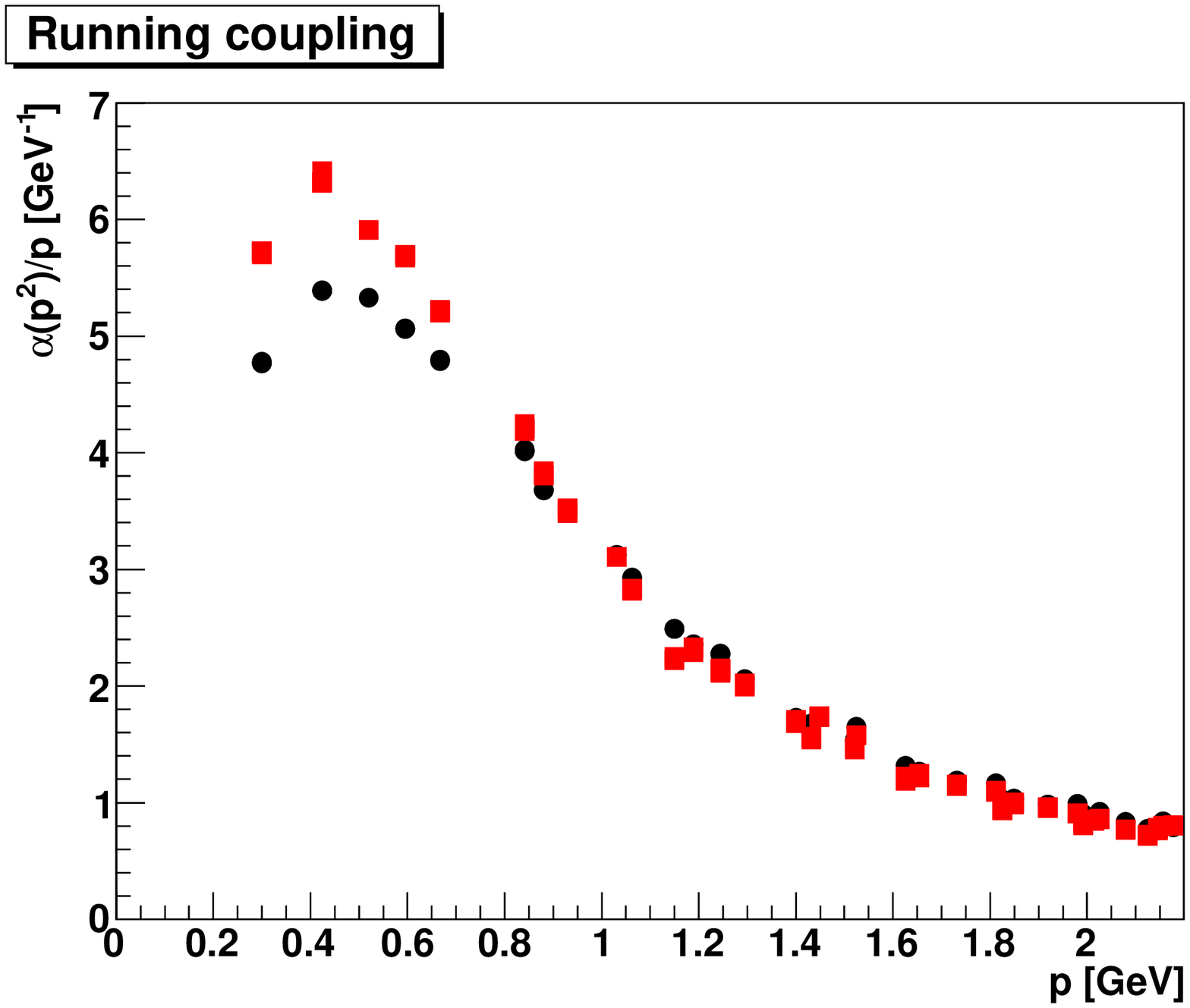}
\caption{\label{fig:gluon}The minimal Landau-gauge gluon propagator $D$ (left panel) and running coupling $\alpha$ (right panel) of G$_2$ Yang-Mills theory compared to SU(3) Yang-Mills theory in three dimensions as a function of momentum $p$, from \cite{Maas:2012wr}.}
\end{figure}

It is thus an interesting question what the effective degrees of freedom are. On the level of the elementary particles, the gluons, no qualitative, and little quantitative difference is found \cite{Maas:2007af,Maas:2010qw}. This also manifests itself in a qualitative similar running coupling, even in the far infrared, see figure \ref{fig:gluon}\footnote{For all results for Yang-Mills theory, the scale has been set by giving the intermediate distance fundamental string-tension a value of (440 MeV)$^2$ \cite{Maas:2007af,Danzer:2008bk}.}. Thus, at the level of gluons, there is no distinct difference.

Another set of effective degrees of freedom often used in Yang-Mills theory are topological ones. Similarly, for G$_2$ Yang-Mills theory vortices \cite{Greensite:2006sm}, monopoles \cite{DiGiacomo:2008nt}, dyons \cite{Diakonov:2010qg}, and instantons \cite{Ilgenfritz:2012aa} have been constructed. Using lattice simulations and cooling, it is indeed possible to verify the existence of topological lumps, which are associated with action lumps and a non-vanishing topological susceptibility of roughly $(150$ MeV$)^4$ \cite{Ilgenfritz:2012aa}, though yet with large systematic errors. Though there exist differences in details, e.\ g.\ vortices are not associated with a  center \cite{Greensite:2006sm}, the salient features of these topological excitations are close to the ones in ordinary SU($N$) Yang-Mills theory. As one can expect from these observations, chiral symmetry is broken in the vacuum in the same way as in ordinary Yang-Mills theory \cite{Danzer:2008bk}.

Thus in total, G$_2$ Yang-Mills theory in the vacuum is very similar to SU($N$) Yang-Mills theories.

\subsection{Finite temperature}

Since the finite-temperature phase transition in SU($N$) Yang-Mills theories is associated with a center-symmetry breaking/restoring phase transition, it was originally anticipated \cite{Holland:2003jy} that there will not be a phase transition in G$_2$ Yang-Mills theory, though the gluonic sector suggested otherwise \cite{Maas:2005ym}. Lattice simulations then indeed found a strong first-order phase transition in G$_2$ Yang-Mills theory \cite{Pepe:2006er,Greensite:2006sm,Cossu:2007dk} using the free energy. However, in practice this is non-trivial due to a bulk transition requiring rather fine lattices \cite{Pepe:2006er,Cossu:2007dk}. This phase transition is also reflected in the behavior of glueballs \cite{Lacroix:2012pt}.

Amazingly, though not being an order parameter, the Polyakov loop also reflected this phase transition. In fact, it is possible to use the Polyakov loops in various representations to describe the phase structure of G$_2$ Yang-Mills theory rather accurately \cite{Wellegehausen:2009rq}. One of the main reasons seems to be that though there is no genuine center symmetry, a distorted three-fold structure is still preserved by G$_2$, which, when breaking the theory down to SU(3), yields the center symmetry, see section \ref{shiggs} below.

This alone is already in remarkable agreement to ordinary Yang-Mills theory. But the similarities are even more pronounced. Since all representations are real, it would have been possible that the chiral transition, as is the case for the adjoint chiral condensate in SU($N$) \cite{Karsch:1998qj,Bilgici:2009jy}, would not show a phase transition or only at a much higher transition temperature. This is not the case, and, within lattice resolution, the chiral condensate shows a response precisely at the same temperature as the Polyakov loop and the free energy \cite{Danzer:2008bk}. As would be naively expected from the comparison to SU($N$) Yang-Mills theory, it is then also found that the topological properties change at the phase transition \cite{Ilgenfritz:2012aa}, especially the topological susceptibility drops.

\begin{figure}
\includegraphics[width=\linewidth]{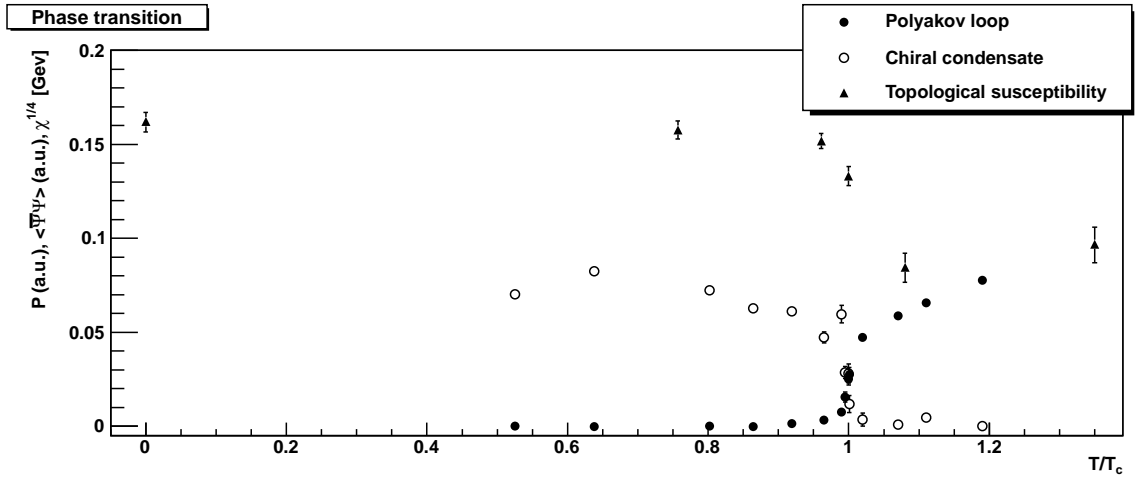}
\caption{\label{fig:phase-ym}The phase diagram of G$_2$ Yang-Mills theory. The critical temperature is taken from \cite{Cossu:2007dk}, the Polyakov loop and chiral condensate from \cite{Danzer:2008bk} and the topological susceptibility from \cite{Ilgenfritz:2012aa}.}
\end{figure}

The resulting phase diagram is shown in figure \ref{fig:phase-ym}. The first order nature is visible, though it requires a detailed study of scaling properties to ascertain it \cite{Cossu:2007dk}. Thus, from the point of view of the phase diagram G$_2$ Yang-Mills theory behaves very similar to the SU(3) case, even though the phase transition is not related to a symmetry. This is one of the reasons why Yang-Mills theory is well suited as a stand-in for QCD thermodynamics, as discussed in section \ref{sqcd-pd}. The reason for the existence of this similarity is besides the approximate three-fold structure \cite{Wellegehausen:2009rq} the fact that the size of the gauge group appears to be more relevant for the phase structure than the center of the group \cite{Pepe:2006er,Holland:2003kg}.

\section{Yang-Mills-Higgs theory}\label{shiggs}

\begin{figure}
\includegraphics[width=\linewidth]{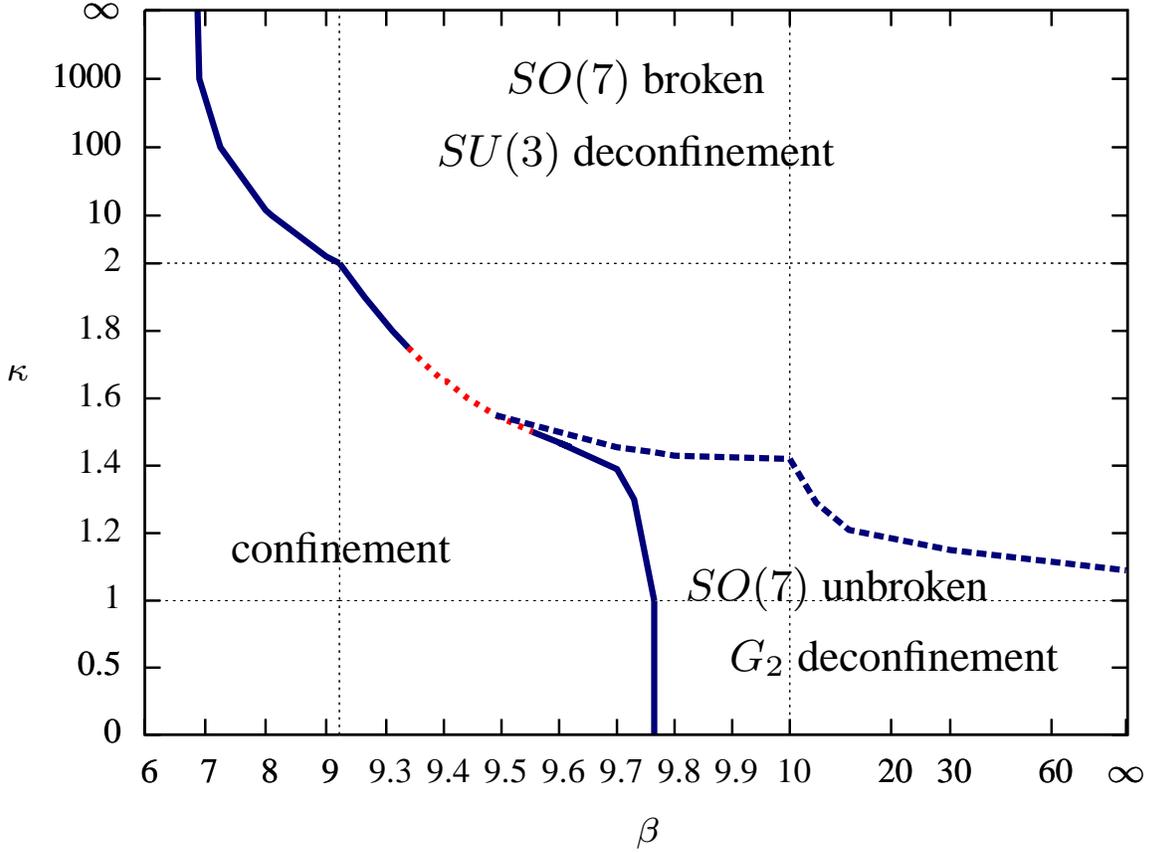}
\caption{\label{fig:phase-higgs}The phase diagram of G$_2$ Yang-Mills-Higgs theory, from \cite{Wellegehausen:2011sc}, as a function of gauge coupling and Higgs hopping parameter at finite temperature.}
\end{figure}

One of the interesting features of G$_2$ is that it has SU(3) as a sub-group. Thus, it appears possible to somehow hide the S$^6$ part of the gauge group using the Higgs mechanism such that just SU(3) remains. In fact, it turns out that a single fundamental Higgs field is sufficient for this purpose \cite{Holland:2003jy,Wellegehausen:2011sc}. In such a more complicated theory it is possible to follow the phase structure at finite temperature, and map a phase diagram in the temperature-Higgs mass plane at infinite four-Higgs coupling \cite{Wellegehausen:2011sc}, as shown in figure \ref{fig:phase-higgs}.

The phase structure is rather intricate at intermediate values of the couplings. Given the large systematic uncertainties encountered in such theories \cite{Bonati:2009pf} a definite answer will remain hard to find. However, this question is highly relevant: If a continuous connection between the SU(3)-like domain and the G$_2$-like domain exists this would have significant implications for the physics of both theories.

The situation becomes much more complicated when introducing (fermionic) matter fields into the theory \cite{Holland:2003jy}. In this case, a hiding with just one Higgs field will inevitably lead to an SU(3) theory with the matter fields in the wrong representation, in particular to real matter fields. Since the natural question is, whether a connection to ordinary QCD is possible, the hiding or breaking  mechanism must complexify the matter fields to lead to the inequivalent fundamental and anti-fundamental representations of QCD. This will likely only be possible, if at all, by manipulating the theory on the level of Weyl fermions, a topic under current investigation \cite{Maas:2012aa}.

\section{G$_2$ QCD}\label{sqcd}

\subsection{Vacuum structure}

When adding $N_f$ fundamental fermions to G$_2$ Yang-Mills theory one arrives at G$_2$ QCD. The vacuum structure of this theory is yet little explored \cite{Holland:2003jy,Maas:2012wr}, but has a number of highly interesting features. The first concerns the spectrum. Due to the group structure, there exists a richer set of color-neutral bound states than in QCD \cite{Holland:2003jy}, both of fermionic and bosonic type. In the boson sector there are as in QCD the glueballs and mesons. In addition, there are also diquarks, since due to the reality of the G$_2$ representations such states are color-neutral, different from QCD, but similar to QC$_2$D. In addition, there are also tetraquarks and heptaquarks consisting out of four and six quarks. Besides these bosonic hadrons there are also fermionic ones. Most notably the hybrid, consisting out of one quark and three gluons, but also a nucleon from three quarks, as well as pentaquarks and heptaquarks from five and seven quarks.

The mass hierarchy of these states will depend strongly on the masses of the quarks, even for degenerate flavors. E.\ g., at heavy quark mass the hybrid will be the lightest particle in the fermionic sector, while the nucleon is expected to take over this role at low quark masses, but will still be heavier than the diquark or mesons. The details of these hierarchy are a dynamical problem.

These bound states are also influenced by the pattern of chiral symmetry breaking. Due to the reality of quarks, G$_2$ has, similarly to QC$_2$D, an enlarged chiral symmetry of U(2$N_f$) \cite{Kogut:2000ek,Holland:2003jy,Hands:2000ei,Maas:2012wr}. This symmetry can be viewed as a flavor symmetry on the level of the Weyl fermions. Of this symmetry an axial U(1) is expected to be broken in the same way as in ordinary QCD by the axial anomaly. Taking for the following a single flavor leaves, in contrast to QCD, still an SU(2) chiral symmetry. This symmetry is spontaneously broken \cite{Maas:2012wr}, like in the quenched case \cite{Danzer:2008bk}, leaving only an U(1) intact. This conserved U(1) can then be associated with a baryon number. The Goldstone bosons of this breaking are then expected to be two diquarks \cite{Maas:2012wr}, just like in QC$_2$D \cite{Strodthoff:2011tz}. These two diquarks represent a flavor-doublet on the level of Weyl fermions.

\begin{figure}
\includegraphics[width=0.5\linewidth]{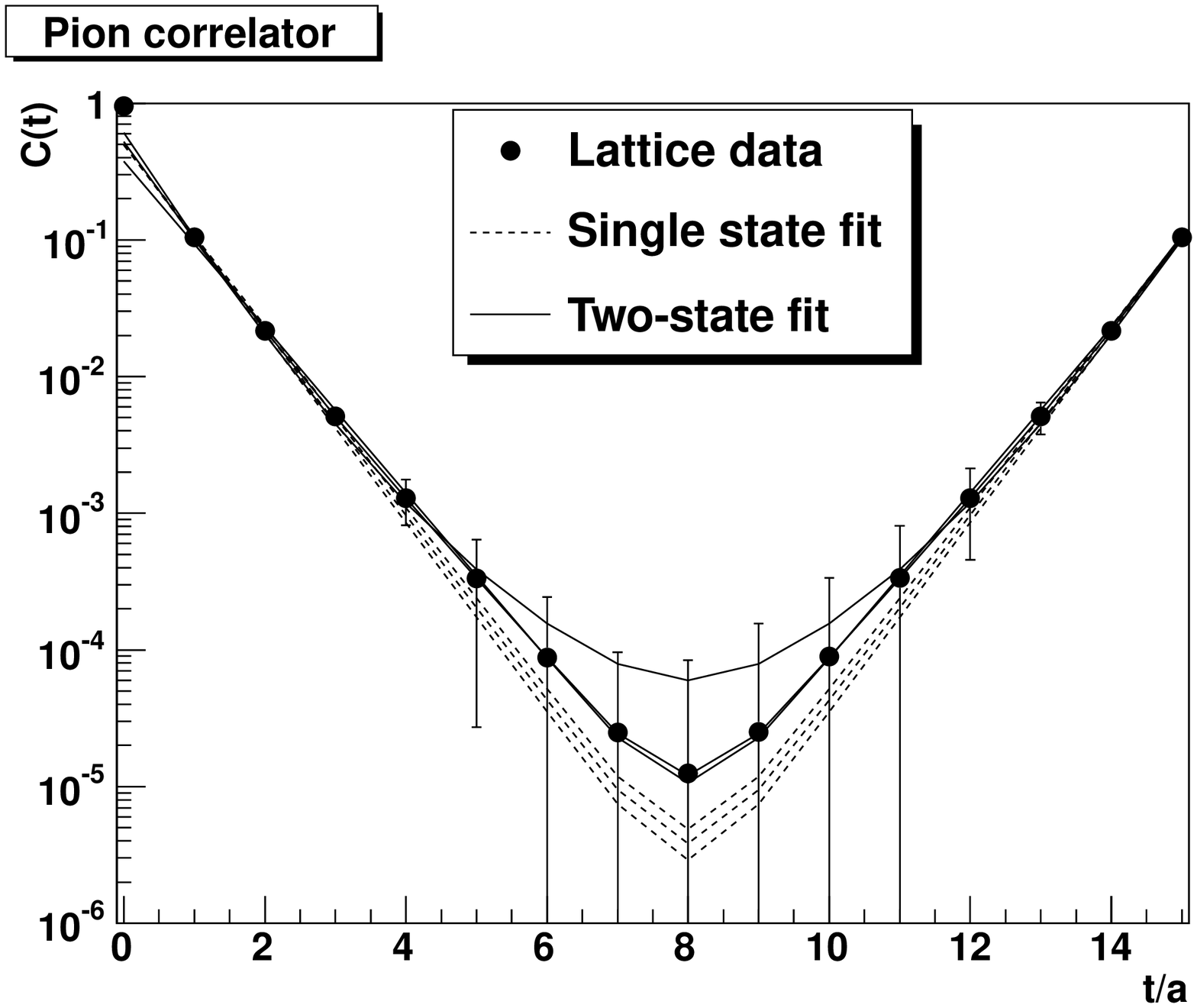}\includegraphics[width=0.5\linewidth]{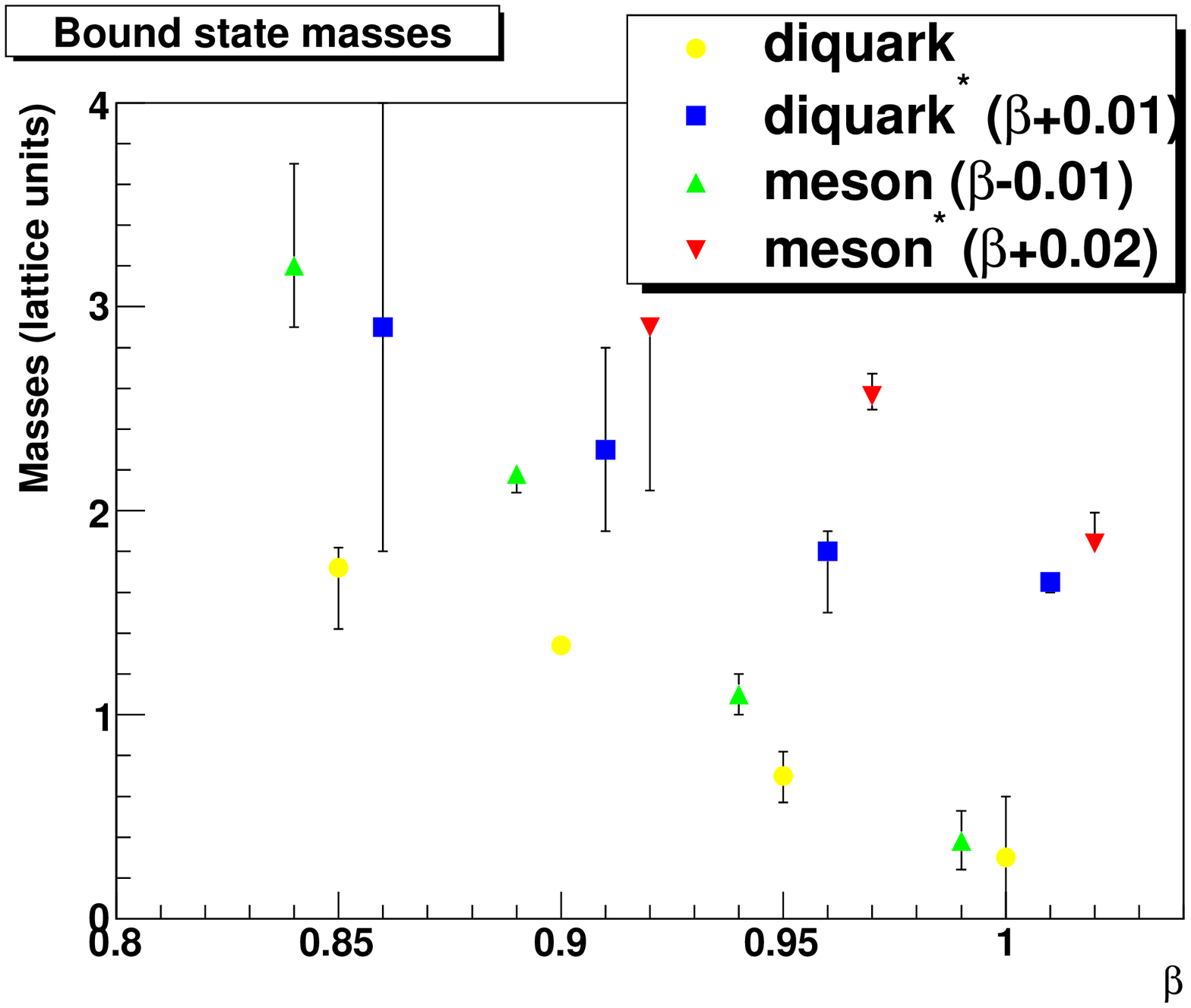}
\caption{\label{fig:mass}The connected part of the diquark/scalar meson correlator with a mass fit (left panel) and the masses for the diquark and the pion as a function of the gauge coupling (right panel), from \cite{Maas:2012wr}. The lattice spacing is strongly-dependent on the lattice parameters.}
\end{figure}

In numerical simulations this is rater hard to identify, as for one Dirac flavor the scalar and the diquarks only differ by disconnected contributions. Furthermore, it turns out that G$_2$ QCD in the range of accessible parameters is very sensitive to both the gauge coupling and the hopping parameter, and has at rather low gauge coupling at fixed hopping parameters already a transition into an unphysical phase \cite{Maas:2012aa}, possibly an Aoki-like phase. Nonetheless, mass determinations are possible, as is demonstrated for $N_f=1$ in figure \ref{fig:mass}. The determination of the vacuum spectrum is thus a challenging task, even at a qualitative level, and an ongoing project \cite{Maas:2012aa}. Especially the mass of the nucleon is relevant, when one turns to the phase diagram.

\subsection{Phase diagram}\label{sqcd-pd}

Due to reality of the representations and the enlarged chiral symmetry, the whole phase diagram for the $N_f=1$ case is both accessible in lattice simulations and relevant. Even besides the fact that G$_2$ QCD is an interesting theory on an intellectual level, there is a number of features which makes it also highly relevant on the level of applications in the continuum limit. First of all, as described in section \ref{sym}, the theory is in the quenched limit very similar to SU($N$) Yang-Mills theories, in contrast to theories with adjoint matter \cite{Karsch:1998qj,Bilgici:2009jy,Hands:2000ei}. Furthermore, the theory has nucleons, and in general fermionic baryons, and thus also nuclei. Hadronic Pauli effects at intermediate densities will thus play a role, in contrast to QC$_2$D \cite{Hands:2006ve,Hands:2011ye,Strodthoff:2011tz,Cotter:2012tt}. No other gauge theory with this combination of features has yet been simulated on a lattice, except without continuum limit \cite{deForcrand:2009dh,Fromm:2011zz}.

This provides the possibility of a number of unprecedented tests of lattice approaches to finite density QCD. It is possible to test explicitly to which extent investigations using analytical continuation in imaginary or isospin chemical potential work (see e.\ g.\ \cite{deForcrand:2010he,Cea:2012ev}), and whether Taylor expansions (see e.\ g.\ \cite{Karsch:2010hm,Borsanyi:2012cr}), Lee-Yang zeros (see e.\ g.\ \cite{Fodor:2004nz}), or other methods (see e.\ g.\ \cite{Fodor:2007vv}) are reliable tools.

Furthermore, and possibly even more important, the G$_2$ QCD lattice phase diagram provides new benchmarks for both models \cite{Leupold:2011zz,Buballa:2003qv,Pawlowski:2010ht} and continuum methods, in the latter case especially functional methods \cite{Pawlowski:2010ht,Braun:2011pp,Maas:2011se}. Furthermore, if breaking G$_2$ QCD to ordinary QCD should be possible, this would be even more helpful, though, of course, at some point the sign problem will prevent a simulation of QCD.

\begin{figure}
\includegraphics[width=\linewidth]{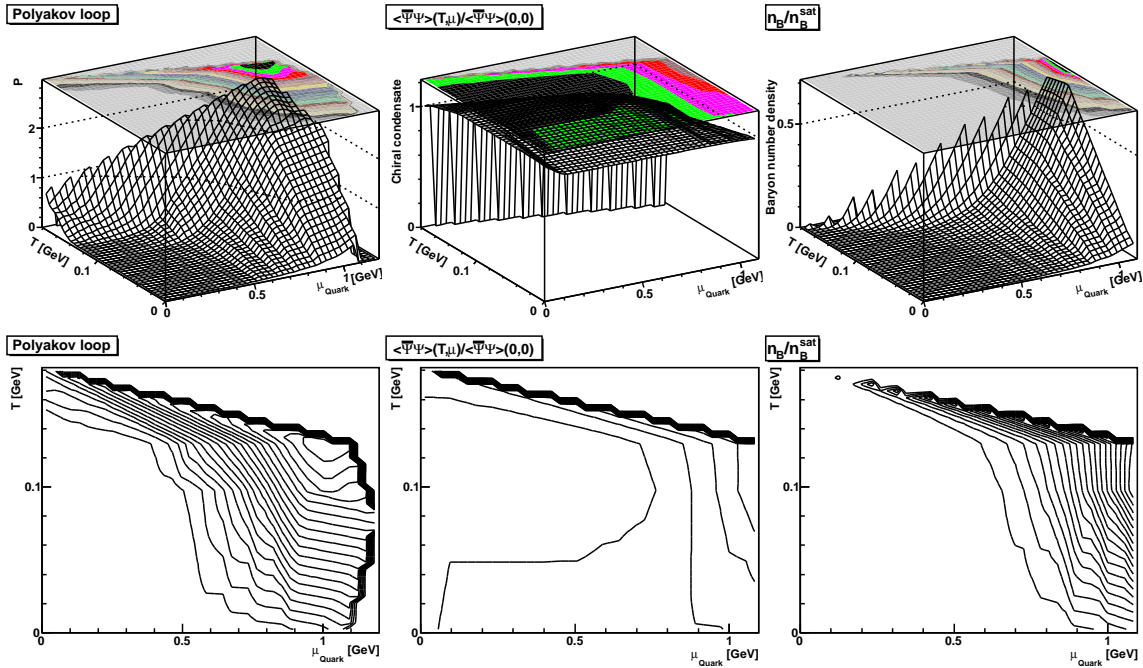}
\caption{\label{fig:pd}The G$_2$ QCD phase diagram for one flavor of quarks. The left panel shows the (unrenormalized) Polyakov loop, the middle panel the normalized chiral condensate, and the right panel the Baryon density, normalized to the saturation density of 14 quarks/lattice site. For details and simulation parameters, see \cite{Maas:2012wr}. Note that at $\mu_{\text{Quark}}\approx 1$ GeV the system starts to become dominated by systematic effects \cite{Maas:2012wr}.}
\end{figure}

The first step in this program is provided by a proof-of-principle showing the accessibility of the phase diagram in lattice calculations, see figure \ref{fig:pd}\footnote{The scale is here chosen to get a zero-density transition temperature of about 160 MeV, and the first excited meson state is used to set the scale scale. This procedure \cite{Maas:2012wr} is strongly affected by systematic errors, and will be improved \cite{Maas:2012aa}.} \cite{Maas:2012wr}. Though so far at a qualitative level, it shows already a structure close to the expected one, including indications \cite{Maas:2012aa} of a silver-blaze point \cite{Cohen:2003kd}, see also section \ref{ssmall}. A more quantitative description will require more detailed calculations, in particular concerning systematic errors \cite{Maas:2012aa}. Nonetheless, the theory shows the low-temperature, low-density ordinary phase, has a transition, likely a cross-over, to a high-temperature phase, and also a transition at finite density. Whether any of these are phase transitions remains to be seen, but so far the finite-density transitions are stronger. Also, first signals of additional structure at zero temperature have been observed \cite{Maas:2012aa}, which may correspond to various phase also observed in QC$_2$D \cite{Strodthoff:2011tz,Cotter:2012tt}. However, more details studies, especially of systematic effects are necessary before definite statements can be made.

\begin{figure}
\includegraphics[width=\linewidth]{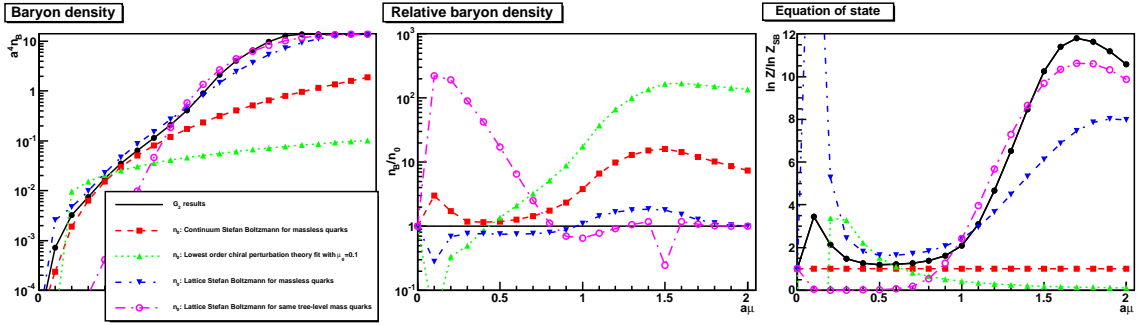}
\caption{\label{fig:eom}The baryon density (left panel) of G$_2$ QCD, compared to that of continuum and lattice \cite{Hands:2006ve} Stefan-Boltzmann results with the same mass or for massless quarks, and to leading order chiral perturbation theory \cite{Hands:2006ve,Kogut:2000ek} with coefficients fitted to the G$_2$ case at intermediate densities. The middle panel shows the corresponding ratios (note the logarithmic scale), and the right-hand panel the integrated equation of state, normalized to the continuum Stefan-Boltzmann case. All results unpublished from \cite{Maas:2012aa}. The value of the lattice constant $a$ is approximately 0.2 fm.}
\end{figure}

Finally, an interesting question is, to which extent the low-temperature case is simple, so that e.\ g., quasi-particle models would be a good description. For this purpose, a comparison to a system of free quarks and to chiral perturbation theory is shown in figure \ref{fig:eom}. While the high-density region, which is dominated by lattice artifacts \cite{Maas:2012wr}, is rather well described by the corresponding free lattice system of quarks, this is not the case at low densities. Here, the equation of state is much more similar to lattice or continuum versions of a gas of free massless quarks, instead of massive ones, though the deviations are still very large at the smallest densities. At the same time, at least leading-order chiral perturbation theory is not able to reproduce even qualitatively the physics of G$_2$ QCD. Thus, non-trivial effects play a dominant role at densities below $a\mu\approx 0.5$, which translates in this case to roughly 500 MeV of quark chemical potential. In this region, highly non-trivial effects have to be dealt with.

\section{Results on a smaller lattice}\label{ssmall}

\begin{figure}[htb]
\begin{center}
\scalebox{1.1}{\includeEPSTEX{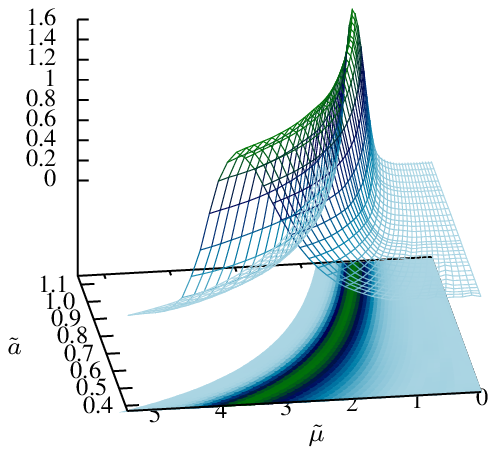}}\hskip10mm
\scalebox{1.1}{\includeEPSTEX{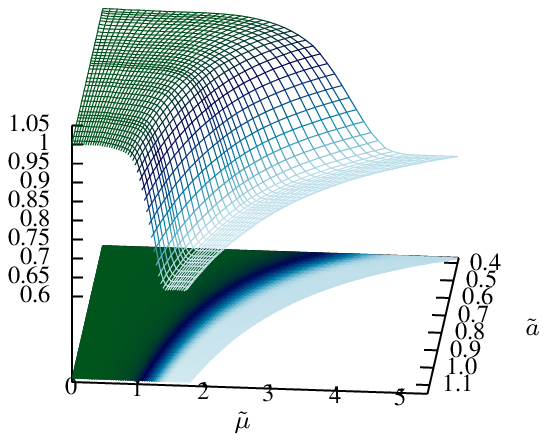}}
\caption{\label{fig:PolChiralSmall}Polyakov loop (left panel) and chiral condensate (right panel) on a $8^3 \times 16$ lattice as a function of chemical potential $\tilde{\mu}$ and lattice spacing $\tilde{a}$. From \cite{Maas:2012aa}.}
\end{center}
\end{figure}

\begin{figure}[htb]
\begin{center}
\scalebox{1.1}{\includeEPSTEX{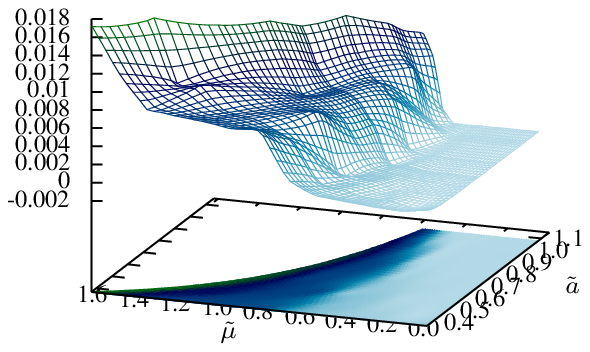}}\hskip10mm
\scalebox{1.1}{\includeEPSTEX{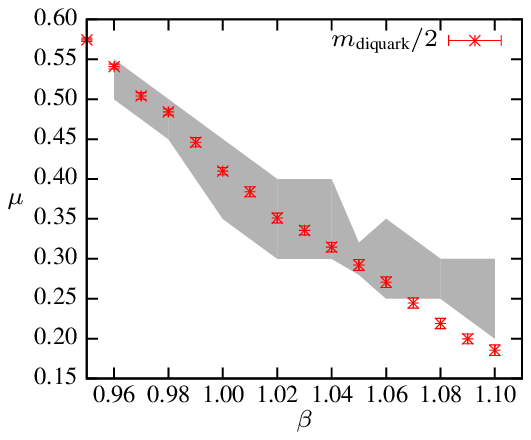}}
\caption{\label{fig:onsetSmall}Quark number density (left panel) and onset transition compared to the diquark mass (right panel) on a $8^3 \times 16$ lattice. From \cite{Maas:2012aa}.}
\end{center}
\end{figure}

Since many of the investigations above are limited by the number of different lattice settings which can be simulated, the use of smaller lattices may help in mapping the phase diagram on a finer grid. However, due to the unphysical bulk transition it is not possible to study the full phase diagram on smaller lattices, especially at finite temperature on lattices with $N_t<5$. 
Nevertheless, at zero temperature $G_2$ QCD is investigated on a $8^3 \times 16$ lattice in
the parameter region $\beta=0.90\ldots1.10$ and $\mu=0\ldots2$. The monopole
density is already sufficiently small, such that the system stays outside the
bulk phase for all values of $\beta$ and $\mu$. 
\begin{figure}[htb]
\begin{center}
\scalebox{1.1}{\includeEPSTEX{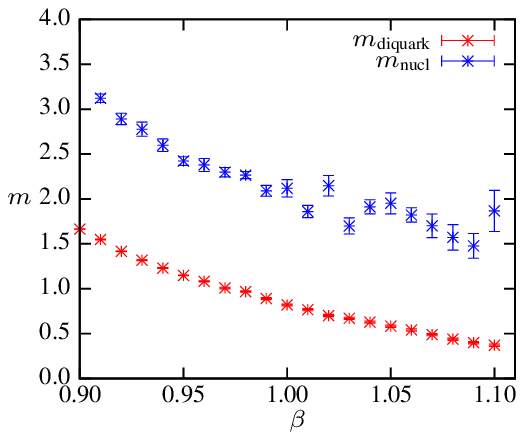}}\hskip10mm
\scalebox{1.1}{\includeEPSTEX{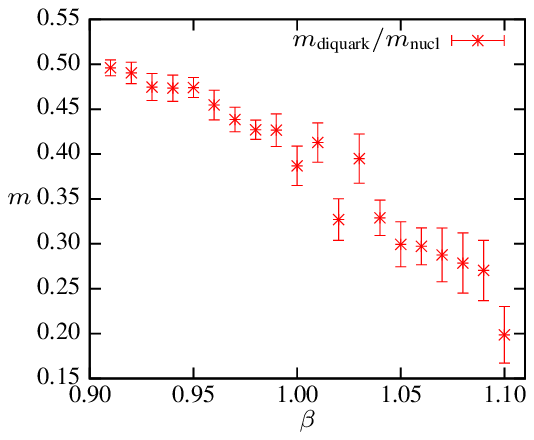}}
\caption{\label{fig:massSmall}Diquark and nucleon mass (left panel) and its ratio (right panel) on a $8^3 \times 16$ lattice. From \cite{Maas:2012aa}.}
\end{center}
\end{figure}
In Fig.~\ref{fig:massSmall} the diquark and nucleon (proton) mass together with its ratio are shown as a function of $\beta$. Assuming a nucleon mass of about $1$ GeV, the diquark mass changes from $\sim 500$ MeV at $\beta=0.90$ to $\sim 200$ MeV at $\beta=1.10$.  On the small lattice the scale is set by
the ground state diquark mass $\tilde{a}(\beta) \equiv m_\text{diquark}(\beta)$. The phase diagram at zero
temperature is then given as a function of the dimensionless parameter $\tilde{\mu}=\mu/m_\text{diquark}$ and the
dimensionless lattice spacing $\tilde{a}$. 
\begin{figure}[htb]
\begin{center}
\scalebox{1.1}{\includeEPSTEX{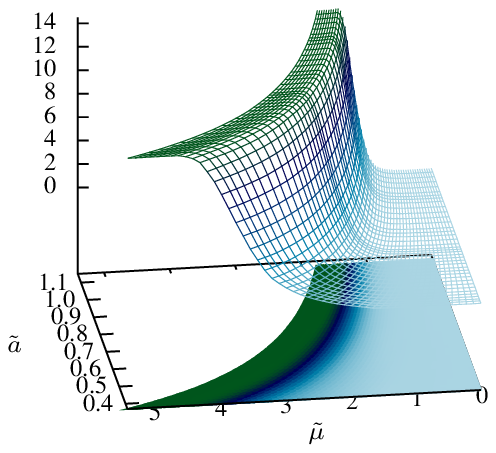}}\hskip10mm
\scalebox{1.1}{\includeEPSTEX{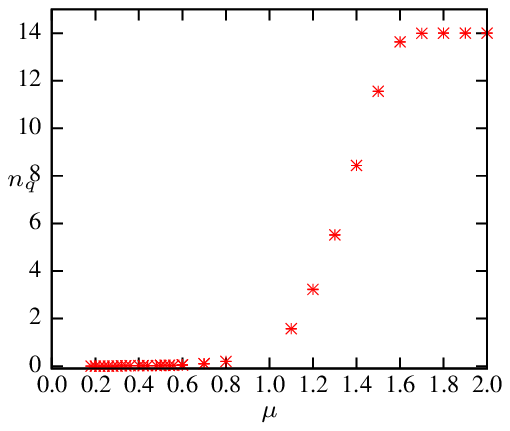}}
\caption{\label{fig:QNDSmall}Quark number density as a function of the lattice spacing $\tilde{a}$ (left panel) and at $\beta=1.05$ (right panel) on a $8^3 \times 16$ lattice. From \cite{Maas:2012aa}.}
\end{center}
\end{figure}
In Fig.~\ref{fig:QNDSmall} the
quark number density is shown. Independent of the lattice spacing the quark number density takes it maximum value of $n_{q, \text{sat}}=2 \cdot N_c \cdot N_f=14$ at large $\tilde{\mu}$. With decreasing lattice spacing, the saturation shifts to larger values of chemical potential, indicating that this saturation is only a lattice artifact. The Polyakov loop and the (renormalized) chiral condensate show almost the same behaviour as on the larger lattices, see Fig.~\ref{fig:PolChiralSmall}.
Furthermore, the onset transition from the vacuum to nuclear matter is studied in
Fig.~\ref{fig:onsetSmall}.
At $\tilde{\mu}_0\approx0.5$, a transition in the quark number density
(left panel) is observed. The value of the onset does almost not depend on the lattice
spacing, indicating that at smaller values of $\tilde{\mu}$ finite size effects
are less important than for larger values of the chemical potential. In the
right panel, the transition (shaded region) is compared to half of the diquark
mass, and a clear coincidence is visible. This indeed verifies that $G_2$ QCD possesses, as advertised above, the
silver blaze property \cite{Cohen:2003kd} for baryon chemical potential, i.e. half of the 
mass of the lightest bound state carrying baryon number is a lower bound for the onset transition to nuclear matter.
With decreasing lattice spacing $\tilde{a}$, a plateau develops for $\tilde{\mu}_0(\tilde{a})<\tilde{\mu}<\tilde{\mu}_1(\tilde{a})$, where the quark
number density is almost constant. For $\tilde{\mu}>\tilde{\mu}_1(\tilde{a})$ it
starts again to increase until it saturates at $\tilde{\mu}_\text{sat}$.

\section{Conclusions}\label{sconclusion}

Concluding, G$_2$ QCD is a highly interesting arena to investigate both conceptual and practical questions. Conceptually, it has already taught us that the center of the gauge group is far less relevant than originally anticipated. Most of the salient features of Yang-Mills theory are also present for this case with trivial center. It can thus be expected that many other questions may be little affected by the center as well. However, it also taught us that the group structure and matter representation is important for the physics. 

Investigating practical applications, which particularly involve benchmarks for models and continuum methods at finite densities and low temperatures, is only a newly emerging field. It has been shown that this is possible. It was furthermore already found that the low and intermediate density regime are quite different from simple systems, confirming the situation in QC$_2$D. To fully control this domain, so important for compact stellar objects, much progress will be needed. G$_2$ QCD will, almost certainly, play an important role in the support of this enterprise in the years to come.

\section{Acknowledgments}

A.\ M.\ is grateful to Julia Danzer, Christof Gattringer, Ernst-Michael Ilgenfritz, and {\v S}tefan Olejn\'ik for the collaboration on these subjects. A.\ M.\ and B.\ W.\ are grateful to Lorenz von Smekal and Andreas Wipf for the collaboration on these subjects, especially on the yet unpublished results \cite{Maas:2012aa} shown in figures \ref{fig:eom}-\ref{fig:onsetSmall}.

\bibliographystyle{bibstyle}
\bibliography{bib}

\end{document}

%% file: potential40.tex
% GNUPLOT: LaTeX picture with Postscript
\begingroup
\scriptsize
  \makeatletter
  \providecommand\color[2][]{%
    \GenericError{(gnuplot) \space\space\space\@spaces}{%
      Package color not loaded in conjunction with
      terminal option `colourtext'%
    }{See the gnuplot documentation for explanation.%
    }{Either use 'blacktext' in gnuplot or load the package
      color.sty in LaTeX.}%
    \renewcommand\color[2][]{}%
  }%
  \providecommand\includegraphics[2][]{%
    \GenericError{(gnuplot) \space\space\space\@spaces}{%
      Package graphicx or graphics not loaded%
    }{See the gnuplot documentation for explanation.%
    }{The gnuplot epslatex terminal needs graphicx.sty or graphics.sty.}%
    \renewcommand\includegraphics[2][]{}%
  }%
  \providecommand\rotatebox[2]{#2}%
  \@ifundefined{ifGPcolor}{%
    \newif\ifGPcolor
    \GPcolortrue
  }{}%
  \@ifundefined{ifGPblacktext}{%
    \newif\ifGPblacktext
    \GPblacktexttrue
  }{}%
  % define a \g@addto@macro without @ in the name:
  \let\gplgaddtomacro\g@addto@macro
  % define empty templates for all commands taking text:
  \gdef\gplbacktext{}%
  \gdef\gplfronttext{}%
  \makeatother
  \ifGPblacktext
    % no textcolor at all
    \def\colorrgb#1{}%
    \def\colorgray#1{}%
  \else
    % gray or color?
    \ifGPcolor
      \def\colorrgb#1{\color[rgb]{#1}}%
      \def\colorgray#1{\color[gray]{#1}}%
      \expandafter\def\csname LTw\endcsname{\color{white}}%
      \expandafter\def\csname LTb\endcsname{\color{black}}%
      \expandafter\def\csname LTa\endcsname{\color{black}}%
      \expandafter\def\csname LT0\endcsname{\color[rgb]{1,0,0}}%
      \expandafter\def\csname LT1\endcsname{\color[rgb]{0,1,0}}%
      \expandafter\def\csname LT2\endcsname{\color[rgb]{0,0,1}}%
      \expandafter\def\csname LT3\endcsname{\color[rgb]{1,0,1}}%
      \expandafter\def\csname LT4\endcsname{\color[rgb]{0,1,1}}%
      \expandafter\def\csname LT5\endcsname{\color[rgb]{1,1,0}}%
      \expandafter\def\csname LT6\endcsname{\color[rgb]{0,0,0}}%
      \expandafter\def\csname LT7\endcsname{\color[rgb]{1,0.3,0}}%
      \expandafter\def\csname LT8\endcsname{\color[rgb]{0.5,0.5,0.5}}%
    \else
      % gray
      \def\colorrgb#1{\color{black}}%
      \def\colorgray#1{\color[gray]{#1}}%
      \expandafter\def\csname LTw\endcsname{\color{white}}%
      \expandafter\def\csname LTb\endcsname{\color{black}}%
      \expandafter\def\csname LTa\endcsname{\color{black}}%
      \expandafter\def\csname LT0\endcsname{\color{black}}%
      \expandafter\def\csname LT1\endcsname{\color{black}}%
      \expandafter\def\csname LT2\endcsname{\color{black}}%
      \expandafter\def\csname LT3\endcsname{\color{black}}%
      \expandafter\def\csname LT4\endcsname{\color{black}}%
      \expandafter\def\csname LT5\endcsname{\color{black}}%
      \expandafter\def\csname LT6\endcsname{\color{black}}%
      \expandafter\def\csname LT7\endcsname{\color{black}}%
      \expandafter\def\csname LT8\endcsname{\color{black}}%
    \fi
  \fi
  \setlength{\unitlength}{0.0500bp}%
  \begin{picture}(3968.00,2976.00)%
    \gplgaddtomacro\gplbacktext{%
      \csname LTb\endcsname%
      \put(600,300){\makebox(0,0)[r]{\strut{}-6.0}}%
      \put(600,681){\makebox(0,0)[r]{\strut{}-4.0}}%
      \put(600,1062){\makebox(0,0)[r]{\strut{}-2.0}}%
      \put(600,1443){\makebox(0,0)[r]{\strut{}0.0}}%
      \put(600,1823){\makebox(0,0)[r]{\strut{}2.0}}%
      \put(600,2204){\makebox(0,0)[r]{\strut{}4.0}}%
      \put(600,2585){\makebox(0,0)[r]{\strut{}6.0}}%
      \put(600,2966){\makebox(0,0)[r]{\strut{}8.0}}%
      \put(660,200){\makebox(0,0){\strut{}0.0}}%
      \put(1320,200){\makebox(0,0){\strut{}0.5}}%
      \put(1981,200){\makebox(0,0){\strut{}1.0}}%
      \put(2641,200){\makebox(0,0){\strut{}1.5}}%
      \put(3302,200){\makebox(0,0){\strut{}2.0}}%
      \put(3962,200){\makebox(0,0){\strut{}2.5}}%
      \put(250,1633){\rotatebox{90}{\makebox(0,0){\strut{}$\tilde{V}_{\rep}/\mu$}}}%
      \put(2311,50){\makebox(0,0){\strut{}$\mu R$}}%
    }%
    \gplgaddtomacro\gplfronttext{%
      \csname LTb\endcsname%
      \put(3479,1541){\makebox(0,0)[r]{\strut{}$\rep=7$}}%
      \csname LTb\endcsname%
      \put(3479,1384){\makebox(0,0)[r]{\strut{}$\rep=14$}}%
      \csname LTb\endcsname%
      \put(3479,1227){\makebox(0,0)[r]{\strut{}$\rep=27$}}%
      \csname LTb\endcsname%
      \put(3479,1070){\makebox(0,0)[r]{\strut{}$\rep=64$}}%
      \csname LTb\endcsname%
      \put(3479,913){\makebox(0,0)[r]{\strut{}$\rep=77$}}%
      \csname LTb\endcsname%
      \put(3479,756){\makebox(0,0)[r]{\strut{}$\rep=77^\prime$}}%
      \csname LTb\endcsname%
      \put(3479,599){\makebox(0,0)[r]{\strut{}$\rep=182$}}%
      \csname LTb\endcsname%
      \put(3479,442){\makebox(0,0)[r]{\strut{}$\rep=189$}}%
    }%
    \gplbacktext
    \put(0,0){\includegraphics{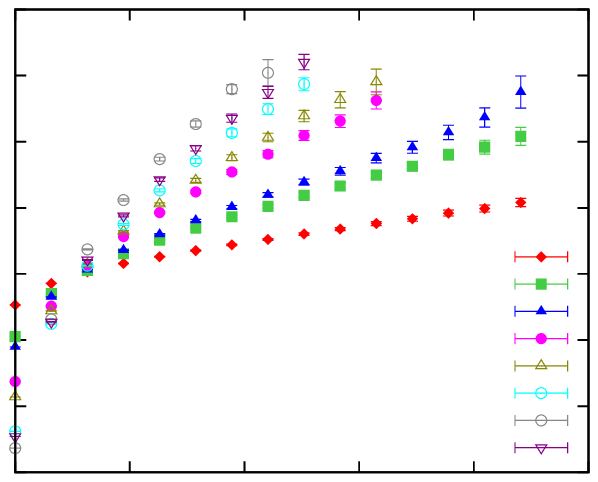}}%
    \gplfronttext
  \end{picture}%
\endgroup

%% file: StringBreakingPot.tex
% GNUPLOT: LaTeX picture with Postscript
\begingroup
\scriptsize
  \makeatletter
  \providecommand\color[2][]{%
    \GenericError{(gnuplot) \space\space\space\@spaces}{%
      Package color not loaded in conjunction with
      terminal option `colourtext'%
    }{See the gnuplot documentation for explanation.%
    }{Either use 'blacktext' in gnuplot or load the package
      color.sty in LaTeX.}%
    \renewcommand\color[2][]{}%
  }%
  \providecommand\includegraphics[2][]{%
    \GenericError{(gnuplot) \space\space\space\@spaces}{%
      Package graphicx or graphics not loaded%
    }{See the gnuplot documentation for explanation.%
    }{The gnuplot epslatex terminal needs graphicx.sty or graphics.sty.}%
    \renewcommand\includegraphics[2][]{}%
  }%
  \providecommand\rotatebox[2]{#2}%
  \@ifundefined{ifGPcolor}{%
    \newif\ifGPcolor
    \GPcolortrue
  }{}%
  \@ifundefined{ifGPblacktext}{%
    \newif\ifGPblacktext
    \GPblacktexttrue
  }{}%
  % define a \g@addto@macro without @ in the name:
  \let\gplgaddtomacro\g@addto@macro
  % define empty templates for all commands taking text:
  \gdef\gplbacktext{}%
  \gdef\gplfronttext{}%
  \makeatother
  \ifGPblacktext
    % no textcolor at all
    \def\colorrgb#1{}%
    \def\colorgray#1{}%
  \else
    % gray or color?
    \ifGPcolor
      \def\colorrgb#1{\color[rgb]{#1}}%
      \def\colorgray#1{\color[gray]{#1}}%
      \expandafter\def\csname LTw\endcsname{\color{white}}%
      \expandafter\def\csname LTb\endcsname{\color{black}}%
      \expandafter\def\csname LTa\endcsname{\color{black}}%
      \expandafter\def\csname LT0\endcsname{\color[rgb]{1,0,0}}%
      \expandafter\def\csname LT1\endcsname{\color[rgb]{0,1,0}}%
      \expandafter\def\csname LT2\endcsname{\color[rgb]{0,0,1}}%
      \expandafter\def\csname LT3\endcsname{\color[rgb]{1,0,1}}%
      \expandafter\def\csname LT4\endcsname{\color[rgb]{0,1,1}}%
      \expandafter\def\csname LT5\endcsname{\color[rgb]{1,1,0}}%
      \expandafter\def\csname LT6\endcsname{\color[rgb]{0,0,0}}%
      \expandafter\def\csname LT7\endcsname{\color[rgb]{1,0.3,0}}%
      \expandafter\def\csname LT8\endcsname{\color[rgb]{0.5,0.5,0.5}}%
    \else
      % gray
      \def\colorrgb#1{\color{black}}%
      \def\colorgray#1{\color[gray]{#1}}%
      \expandafter\def\csname LTw\endcsname{\color{white}}%
      \expandafter\def\csname LTb\endcsname{\color{black}}%
      \expandafter\def\csname LTa\endcsname{\color{black}}%
      \expandafter\def\csname LT0\endcsname{\color{black}}%
      \expandafter\def\csname LT1\endcsname{\color{black}}%
      \expandafter\def\csname LT2\endcsname{\color{black}}%
      \expandafter\def\csname LT3\endcsname{\color{black}}%
      \expandafter\def\csname LT4\endcsname{\color{black}}%
      \expandafter\def\csname LT5\endcsname{\color{black}}%
      \expandafter\def\csname LT6\endcsname{\color{black}}%
      \expandafter\def\csname LT7\endcsname{\color{black}}%
      \expandafter\def\csname LT8\endcsname{\color{black}}%
    \fi
  \fi
  \setlength{\unitlength}{0.0500bp}%
  \begin{picture}(3968.00,2976.00)%
    \gplgaddtomacro\gplbacktext{%
      \csname LTb\endcsname%
      \put(600,300){\makebox(0,0)[r]{\strut{}-2}}%
      \put(600,596){\makebox(0,0)[r]{\strut{}-1}}%
      \put(600,892){\makebox(0,0)[r]{\strut{} 0}}%
      \put(600,1189){\makebox(0,0)[r]{\strut{} 1}}%
      \put(600,1485){\makebox(0,0)[r]{\strut{} 2}}%
      \put(600,1781){\makebox(0,0)[r]{\strut{} 3}}%
      \put(600,2077){\makebox(0,0)[r]{\strut{} 4}}%
      \put(600,2374){\makebox(0,0)[r]{\strut{} 5}}%
      \put(600,2670){\makebox(0,0)[r]{\strut{} 6}}%
      \put(600,2966){\makebox(0,0)[r]{\strut{} 7}}%
      \put(660,200){\makebox(0,0){\strut{} 0}}%
      \put(1210,200){\makebox(0,0){\strut{} 1}}%
      \put(1761,200){\makebox(0,0){\strut{} 2}}%
      \put(2311,200){\makebox(0,0){\strut{} 3}}%
      \put(2861,200){\makebox(0,0){\strut{} 4}}%
      \put(3412,200){\makebox(0,0){\strut{} 5}}%
      \put(3962,200){\makebox(0,0){\strut{} 6}}%
      \put(370,1633){\rotatebox{90}{\makebox(0,0){\strut{}$\tilde{V}_{\rep}/\mu$}}}%
      \put(2311,50){\makebox(0,0){\strut{}$\mu R$}}%
    }%
    \gplgaddtomacro\gplfronttext{%
    }%
    \gplbacktext
    \put(0,0){\includegraphics{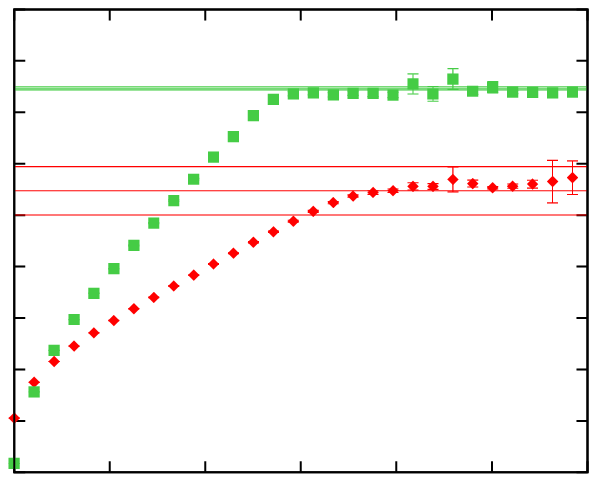}}%
    \gplfronttext
  \end{picture}%
\endgroup